\documentclass[twocolumn,showpacs,preprintnumbers,amsmath,amssymb,aps,prb]{revtex4}

\usepackage{graphicx}% Include figure files
\usepackage{dcolumn}% Align table columns on decimal point
\usepackage{bm}% bold math
\usepackage{subfigure}
\usepackage{natbib}
\usepackage{multirow}

\begin{document}

\title{Time resolved photoemission of Sr$_2$IrO$_4$\\}

\author{C. Piovera$^{1}$, V. Brouet$^{2}$, E. Papalazarou$^{2}$, M. Caputo$^{2}$, M. Marsi$^{2}$, A. Taleb-Ibrahimi${^3}$, B. J. Kim$^4$, and L. Perfetti$^{1}$}

\affiliation{
$^{1}$ Laboratoire des Solides Irradi\'{e}s, Ecole Polytechnique, CNRS, CEA, Universit\'e Paris-Saclay, 91128 Palaiseau, France}

\affiliation{
$^{2}$ Laboratoire de Physique des Solides, CNRS, Univ. Paris-Sud, Université Paris-Saclay, 91405 Orsay Cedex, France}

\affiliation{
$^{3}$ Synchrotron SOLEIL, L'Orme des Merisiers, Saint-Aubin-BP 48, F-91192 Gif sur Yvette, France}

\affiliation{
$^{4}$ Max Planck Institute for Solid State Research, Heisenbergstraße 1, D-70569 Stuttgart, Germany}

\date{\today}

\begin{abstract}
We investigate the temporal evolution of electronic states in strontium iridate Sr$_2$IrO$_4$. The time resolved photoemission spectra of intrinsic, electron doped and the hole doped samples are monitored in identical experimental conditions. Our data on intrinsic and electron doped samples, show that doublon-holon pairs relax near to the chemical potential on a timescale shorter than $70$ fs. The subsequent cooling of low energy excitations takes place in two step: a rapid dynamics of $\cong120$ fs is followed by a slower decay of $\cong 1$ ps. The reported timescales endorse the analogies between Sr$_2$IrO$_4$ and copper oxides.

\end{abstract}


\maketitle

The layered Sr$_2$IrO$_4$ is an ideal system where to explore electronic correlations, electron-phonon coupling and antiferromagnetic ordering. Strontium iridate is a quasi-two-dimensional compound with partially filled 5d shell and moderate Coulomb repulsion. In spite of the large extension of 5d orbitals, the subtle interplay of spin-orbit interaction, crystal field splitting and antiferromagnetic interaction leads to an insulating state \cite{Kim_ARPES,Cyril}. Due to super-exchange interaction, the groundstate of Sr$_2$IrO$_4$ holds canted antiferromagnetic ordering of the spins on a squared lattice \cite{Kim_RXS}. As in the case of copper oxides\cite{Millis}, the paramagnetic insulator can be viewed as an intermediate Mott-Slater system that is stabilized by short-range correlations of the antiferromagnetic order \cite{Watanabe}. The analogy with cuprates can be pushed further \cite{Kim_Doped,Baumberger}, insofar as doped Sr$_2$IrO$_4$ is considered a promising candidate where to observe high temperature superconductivity \cite{Kim_Gap}.

It is worth questioning whether iridates and cuprates display the same dynamical behavior upon photoexcitation. In this respect, the insulating copper oxides have already been characterized by exhaustive experiments of transient absorption\cite{Okamoto}. H. Okamoto \emph{et al.} measured at different probing frequencies, thereby disentangling the Drude component from the mid-gap response. It follows that mid-gap states arise on a timescale of 40 fs and experience an initial decay within 200 fs. The iridates seem to display a similar response, although the reported experiments have been performed with probing energy exceeding the optical gap value\cite{Gedikkata,Gedikkatona}. D. Hsieh \emph{et al.} observed a biexponential kinetic and analyzed the effects of the magnetic transition on the relaxation time \cite{Gedikkata}.

The purpose of this work is to directly follow the relaxation of electronic states in Sr$_2$IrO$_4$. We report a time resolved photoemission experiment of the intrinsic and chemically doped samples. Our data indicate that doublon-holon pairs are highly unstable against the formation of mid-gap states. The dynamics of such emerging excitations follow a biexponential law that is compatible to the transient absorption reported in La$_2$CuO$_4$ \cite{Okamoto}. Our results provide further insights on the ultrafast electron relaxation in quasi-two-dimensional Mott insulators with strong entiferromagnetic coupling and reinforce the analogies between iridates and cuprates.

\emph{Methods}: Samples have been synthesized by a self flux method \cite{Kim_RXS}. Electron doping is obtained by substituting 3\% of the Sr atoms with La, whereas hole doping is done by replacing 15\% of Sr atoms with Rh. The crystals have been characterized by X-ray diffraction, resistivity, magnetization measurements and angle resolved photoelectron spectroscopy. Time resolved photoemission experiments were performed on the FemtoARPES setup \cite{FemtoARPES}, using a Ti:sapphire laser system delivering 35 fs pulses at 1.55 eV (780 nm) with 250 kHz repetition rate. Part of the laser beam is used to generate 6.3 eV photons through cascade frequency mixing in BBO crystals. The 1.55 eV and 6.3 eV beams are employed to photoexcite the sample and induce photoemission, respectively. The energy resolution is $\cong$ 60 meV (limited by the energy bandwidth of the UV pulses) and the temporal resolution is $\cong$ 60 fs. All the time resolved photoemission measurements were performed at room temperature and at the base pressure of $7\times10^{-11}$ mbars.

\begin{figure}
\begin{center}
\includegraphics[width=1\columnwidth]{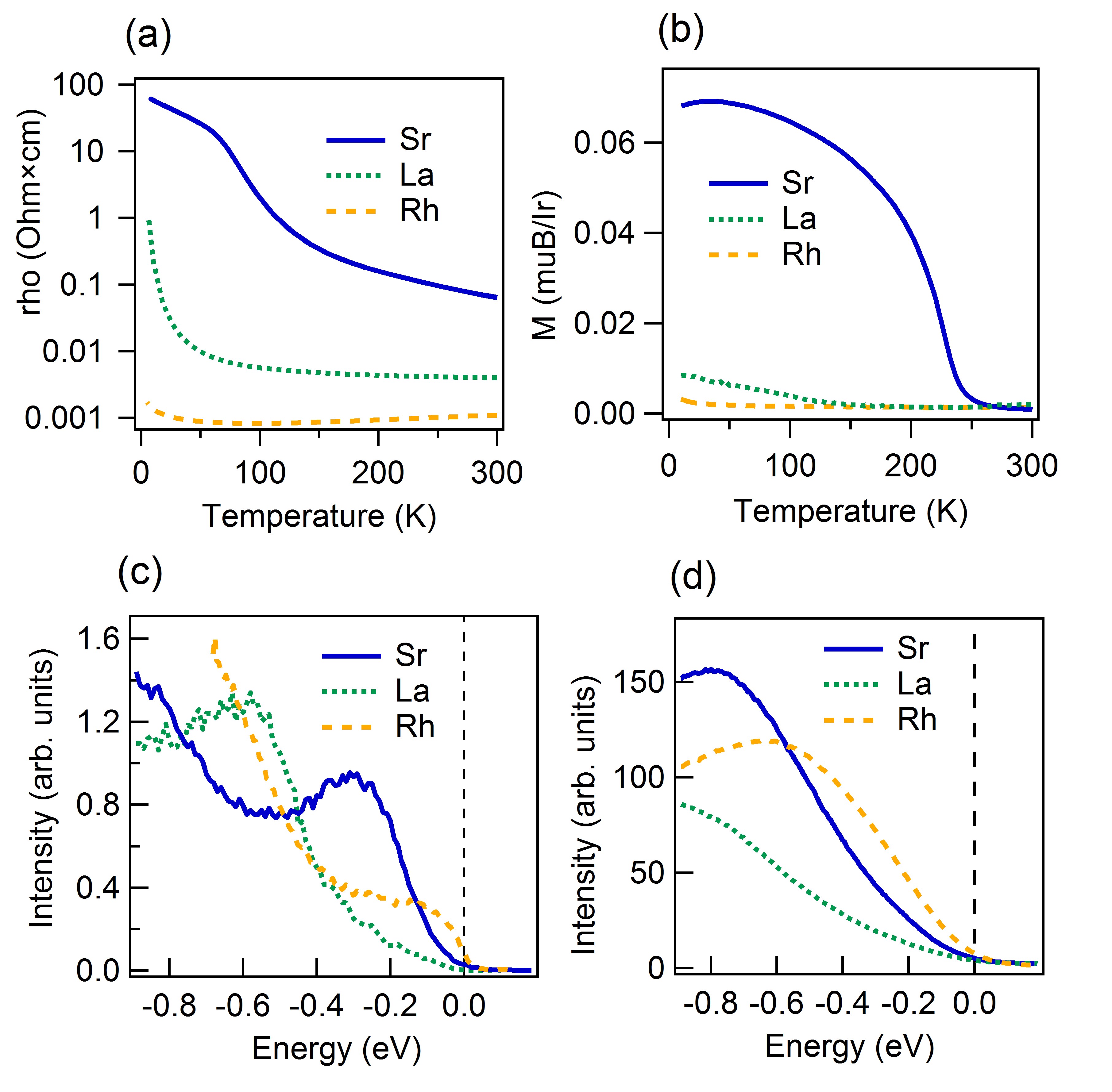}
\caption{In plane resistivity (a) and magnetization curves (b) of intrinsic and doped Sr$_2$IrO$_4$. EDCs acquired at the $X$ point of the Brillouin zone with photon energy of 70 eV (c) and acquired at center of the Brillouin zone with photon energy of 6.3 eV (d). The intrinsic (solid line), electron-doped (dotted line) and hole-doped (dashed line) compound are marked by the notations Sr, La and Rh, respectively.}
\label{Fig1}
\end{center}
\end{figure}

\emph{Samples Characterization}: Figure \ref{Fig1}(a,b) show examples of the resistivity and magnetization measurements in three different samples. The intrinsic Sr$_2$IrO$_4$ exhibits an insulating behavior with an activation energy of 30-70 meV. Upon chemical doping, the conductivity drops by two orders of magnitude. The magnetic phase transition is barely visible at electron-doping (3\% La) and it vanishes at hole-doping (15\% Rh).

According to the literature, the lower Hubbard band has orbital  character $J_{eff}=1/2$ and reaches the nearest distance from the chemical potential at the $X$ point of the Brillouin zone \cite{Kim_ARPES,Brouet}. As shown in Fig. \ref{Fig1}(c) the Hubbard peak is at $\cong-0.25$ eV in the intrinsic sample and shifts to $\cong-0.6$ eV after La substitution \cite{Brouet}. Upon hole doping (15\% Rh), the peak of the $J_{eff}=1/2$ band moves at $\cong-0.1$ eV from the chemical potential, giving rise to a small pseudogap instead of a quasiparticle crossing. As explained in our recent ARPES work \cite{Brouet}, we can estimate a Mott gap of $0.6-0.7$ eV from the difference of the $J_{eff}=1/2$ position in hole doped and electron doped compound. This value is consistent with Scanning Tunneling Spectroscopy (STS) measurements on sample regions with no defects \cite{Okada,Dai}.
The presence of defects is an unavoidable complexity that has strong impact on the low energy physics of iridates \cite{Brouet,Okada,Dai,Glamazda}. %Indeed, Fig. \ref{Fig1}(c) suggests that a large density of mid-gap states are present both in the intrinsic and the hole doped compound. 
STS experiments confirmed that defects induce a local collapse of the Mott gap \cite{Okada} and established a relation between mid-gap states and oxygen impurities \cite{Dai}.

Finally, we show in Fig. \ref{Fig1}(d) the Energy Distribution Curves (EDC) acquired with 6.3 eV photon energy at the center of the Brillouin zone and normalized to the photon flux. In the intrinsic sample, the direct photoemission from the $J_{eff}=1/2$ band and the weak unklapp of the $J_{eff}=3/2$ band should peak at electron energy $-1.2$ eV and $-0.5$ eV respectively \cite{Kim_ARPES,Brouet}. However, the signal in Fig. \ref{Fig1}(d) mainly arises from photoelectronic emission assisted by surface roughness and phonons. Accordingly, we verified that EDCs acquired with 6.3 eV photon energy are dispersionless when scanning the emission angle from 0 to 30 degrees. It is reasonable to assume that the EDCs of  Fig. \ref{Fig1}(d) provide a rough indication of the electron-removal spectral function integrated over the wavevector index. In agreement with this assumption, we observe a large shift of spectral weight upon electron or hole doping.

\begin{figure} \begin{center} \includegraphics[width=0.95\columnwidth]{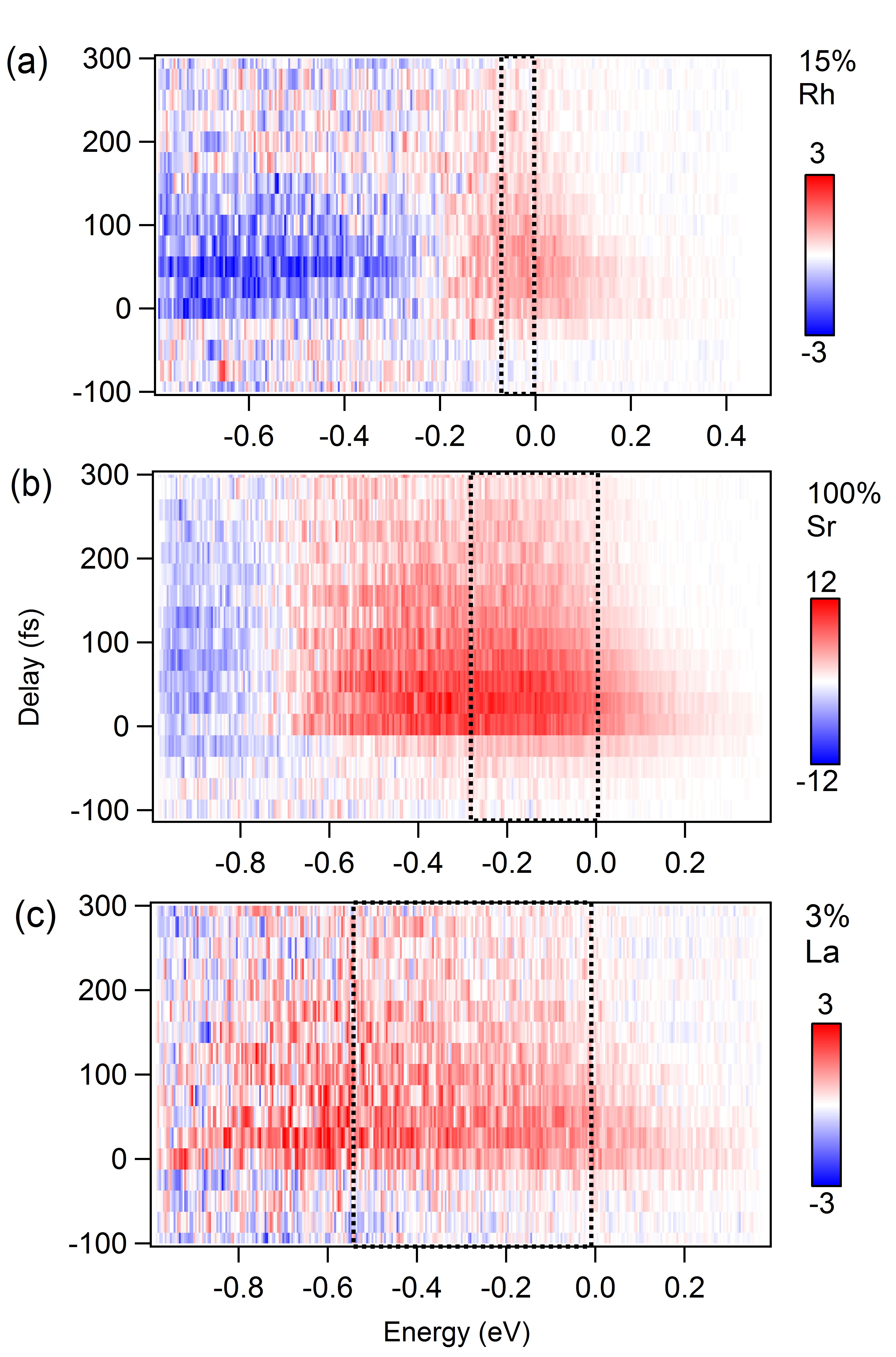} \caption{Map of the pump-on minus pump-off signal acquired with 6.3 eV photons as a function of energy and pump-probe delay. The intensity is normalized with respect to the photon flux and the pump fluence has been set to $\cong 0.7$ mJ/cm$^2$. Data on Rh substituted, intrinsic and La substituted sample are shown in panel (a), (b) and (c) respectively. The dotted rectangles stand for the energy window comprised between the lower Hubbard band and the chemical potential. } \label{Fig2} \end{center} \end{figure}

\emph{Properties of the photoexcited state}:
Exact diagonalisation calculation  indicate that the $J_{eff}=1/2$ electron-hole continuum covers the interval 0.5-1.5 eV \cite{Min}. One may expect that upon pumping with 1.55 eV photons, the photoexicted electrons would accumulate in the form of doublon at the bottom of the upper Hubbard band. However, this scenario is in sharp contrast with our experimental results. Figure \ref{Fig2} reports the pump-on minus pump-off EDCs acquired on different samples as a function of delay time. The differential intensity is plotted on a colorscale where blue and red stand for photoinduced reduction and increase of photoelectron yield, respectively. Intensity map of Fig. \ref{Fig2}(a), \ref{Fig2}(b) and \ref{Fig2}(c) have been acquired on Rh-substituted, intrinsic and La-substituted compound. We employed a pumping fluence of 0.7mJ/cm$^2$, leading to $\sim 0.04$ excitations per iridium atom. The data indicate that photoexcited electrons do not accumulate in the upper Hubbard band but relax in electronic states near to the chemical potential. This finding is in contrast to recent time resolved photoemission measurements of the Mott insulator UO$_2$ \cite{UO2}. We recall that Sr$_2$IrO$_4$ and UO$_2$ differ in terms of dimensionality, gap size and impact of the antiferromagnetic correlations. Therefore, two possible scenarios may explain the distinct electron dynamics in these compounds. First, the Sr$_2$IrO$_4$ has a Mott gap (0.6 eV) much smaller than the one of UO$_2$ (2.3 eV) \cite{UO2}. Therefore the rate of multi-phonon and multi-magnon emission  \cite{Lenarcic1,Lenarcic2} could be strong enough to relax electrons across the correlation gap of Sr$_2$IrO$_4$ whereas it may be negligible in UO$_2$. Second, we know from STS data \cite{Okada,Dai} that defects locally disrupt the narrow Mott gap of Sr$_2$IrO$_4$. As a consequence, the excited electrons may find viable paths to relax from the upper Hubbard band down to lower lying energies.

We outline in Fig. \ref{Fig2}(a-c) that a photoinduced increase of photoemission yield extends well below the chemical potential. A combination of intrinsic and extrinsic effects can explain this finding. On one hand, an intrinsic increase of spectral weight between the lower Hubbard band and the chemical potential (dotted rectangles in Fig. \ref{Fig2}) is expected because of the partial filling of the Mott gap. This behavior has been observed in 1T-TaS$_2$ at comparable excitation densities and should be a general property of Mott insulators with narrow gap \cite{TaS2,TaS2_bis}. On the other hand, an extrinsic and time dependent shift of the EDC can arise from the sudden change of dielectric properties at the surface of the sample. We already observed spectral shift generated by local fields in copper oxides \cite{Piovera}, small gap semiconductors \cite{Mauchain} and semimetals \cite{Papalazarou}. As in the case of surface photovoltage, we expect the energy displacement to be more important in the intrinsic compound than in samples with Rh or La substitutions. Future experiments with high harmonic sources \cite{UO2,Rossnagel} could access the $X$ point of the Brillouin zone and may shed light on this issue.

\begin{figure}
\begin{center}
\includegraphics[width=1\columnwidth]{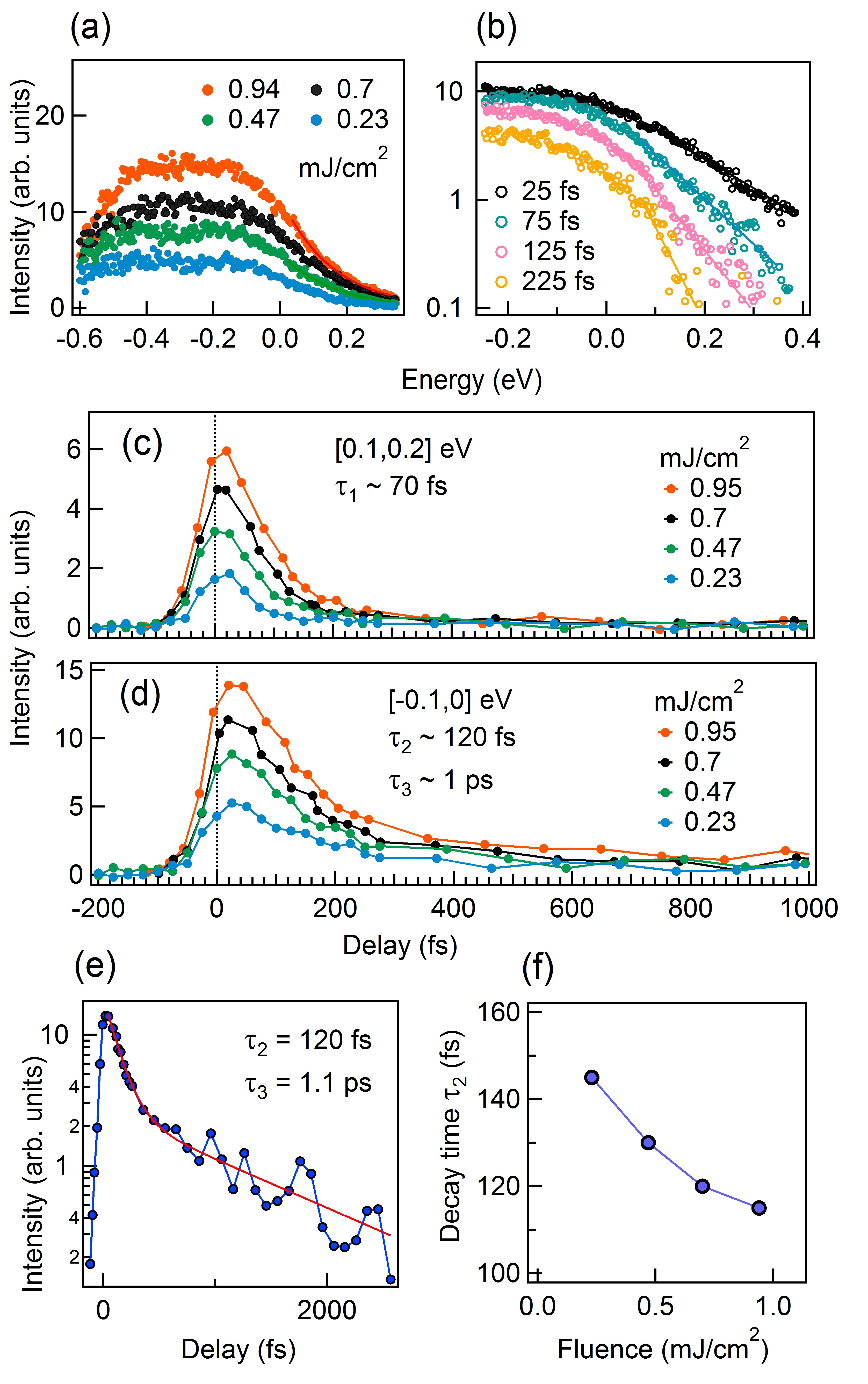}
\caption{All data of this figure refer to the intrinsic sample. (a) Energy distribution of the pump-on minus pump-off signal acquired at delay time of 25 fs at different pumping fluence. (b) Energy distribution of the pump-on minus pump of signal acquired with 0.7 mJ/cm$^2$ at different pump-probe delays. The solid lines stands for the exponential fit of the spectral tail at energy $\geq 0.1$ eV. Evolution of the signal integrated in the energy interval $[0.1,0.2]$ eV (panel (c)) and $[-0.1,0]$ (panel (d)) as a function of delay time and for different fluences. e) Logarithmic plot of the $[-0.1,0]$ eV signal for pump fluence of 0.7 mJ/cm$^2$. f) Dependence of the decay time $\tau_2$ on the pump fluence.}
\label{Fig3}
\end{center}
\end{figure}

\emph{Dynamics of the electrons:} Figure \ref{Fig3}(a) shows pump-on minus pump-off EDCs acquired in the intrinsic sample, at delay time of 25 fs and different pumping fluences. The curves have similar shape and are nearly proportional to the pumping fluence. Therefore, the photoexcited state is still far from the saturation regime of a collapsed Mott insulator \cite{TaS2_bis}.
We show in Fig. \ref{Fig3}(b) the pump-on minus pump-off EDCs of the intrinsic sample acquired with excitation fluence of 0.7 mJ/cm$^2$ at different pump-probe delays. Notice that electrons at excitation energy higher than $\cong0.1$ eV follow an exponential distribution $\exp(-\epsilon/kT_e)$ with temperature scale $T_e$ attaining a maximal value of $1900$ K. In order to gain further insights on the electronic relaxation, we plot in Fig. \ref{Fig3}(c) the temporal evolution of the signal integrated in the energy interval [0.1,0.2] eV. This spectral region is below the upper Hubbard band and above the chemical potential. Therefore the transient signal recorded in [0.1,0.2] eV is an excellent indicator of the energy flow from doublon-holons pairs into lower energy excitations. A fit accounting for the cross-correlation between pump and probe beam provides the decay time $\tau_1\cong$ 70 fs. This timescale does not change with respect to photoexcitation density and indicates that doublon-holons pairs relax near the chemical potential on an very short timescale. Our results are in line with the sudden decay of the Drude response observed by Okamoto \emph{et al.} in photoexcited cuprates \cite{Okamoto}.

Next we turn on the temporal evolution of the pump-probe signal in the spectral region where mid-gap states accumulate. We cannot resolve any finite rise time in the pump-probe signal, indicating that mid-gap states are formed within less than 60 fs. Figure \ref{Fig3}(d) shows the evolution of the photoelectron intensity integrated in the spectral range $[-0.1,0]$ eV. The relaxation of the mid-gap signal follows a biexponential decay with time constant $\tau_2\cong120$ fs and $\tau_3\cong1$ ps. As shown in Fig. \ref{Fig3}(e)), the presence of the two timescales is clearly resolved by plotting the decay curve on a logarithmic scale. Similar dynamics have been also reported in experiments of transient reflectivity at 1.5 eV \cite{Gedikkata}. Most importantly, we outline here the clear analogy between iridates and cuprates. According to Okamoto \emph{et al.}, the midgap states of photoexcited La$_2$CuO$_4$  display an initial relaxation taking place within 200 fs \cite{Okamoto}. Such fast recovery of the charge gap is typical of quasi-two-dymensional Mott insulators with strong antiferromagnetic correlations, whereas does not take place in correlated insulators where the partial gap filling comes along with large structural distortions \cite{LaVS3,VO2}.

\begin{figure}
\begin{center}
\includegraphics[width=1\columnwidth]{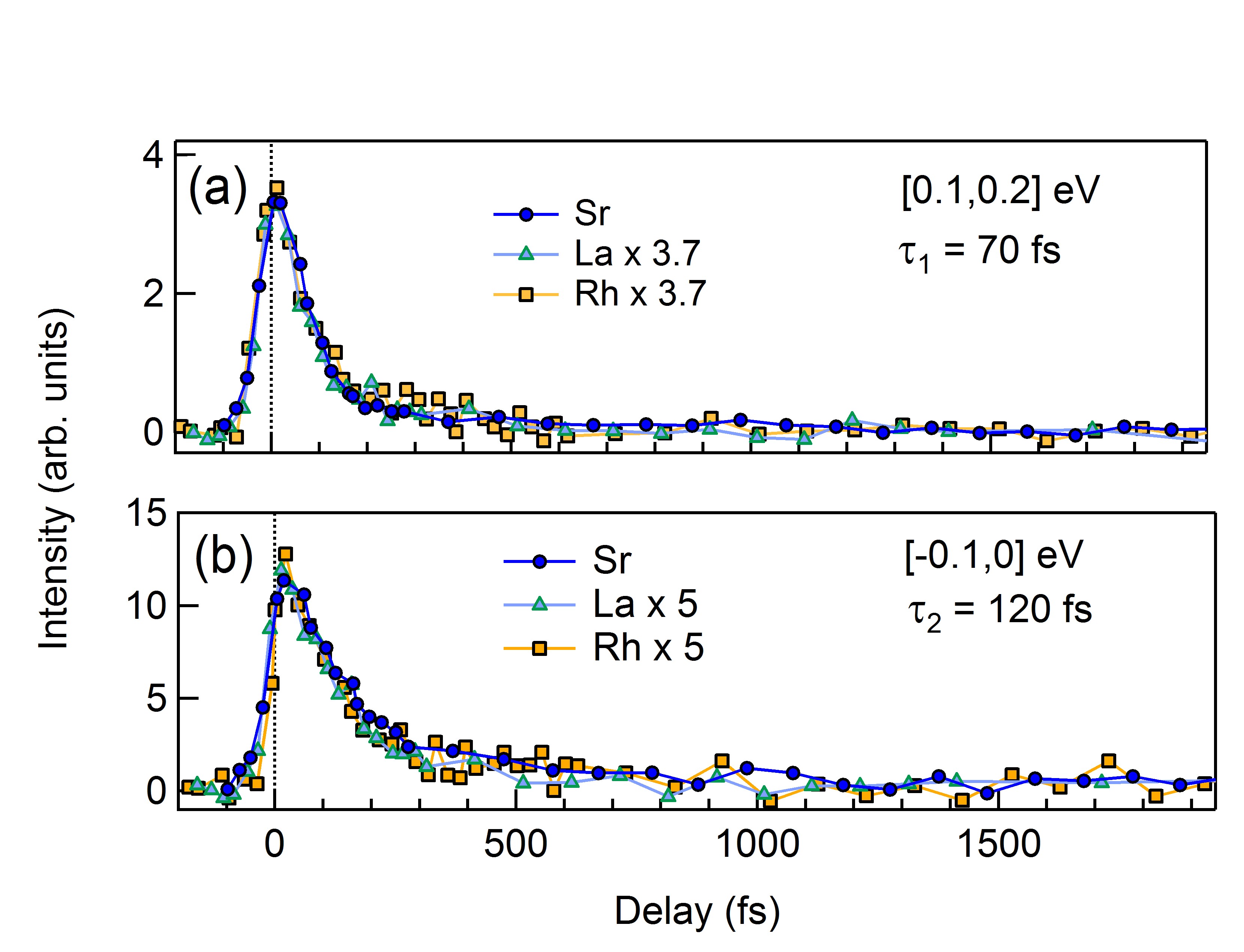}
\caption{Evolution of the pump-probe signal integrated in the energy interval $[0.1,0.2]$ eV (panel (a)) and $[-0.1,0]$ (panel (b)). The circles, triangles and squares stand for the intrinsic, La substituted and Rh substituted sample.}
\label{Fig4}
\end{center}
\end{figure}

As in the case of cuprates \cite{Okamoto,BSCCO,Piovera}, we ascribe the $\tau_2$ decay to the energy dissipation via emission of optical phonons or localized vibrations. The slower dynamics $\tau_3$ is instead due to anharmonicity and acoustic phonon emission. Within our experimental accuracy we could not observe any fluence dependence in the slow timescale $\tau_3$. Conversely, Fig. \ref{Fig3}(f) shows a weak increase of $\tau_2$ by lowering the photoexcitation density.  At last, Fig. \ref{Fig4} compares the dynamics of the signal acquired on the intrinsic, Rh-substitued and La-substituted sample. Strikingly, we observe an identical temporal behavior in the three different samples. Independently on the doping, the excited electrons integrated in the interval [0.1,0.2] eV decay with time constant $\tau_1=70$ fs. Instead, the midgap states in the energy window [-0.1,0] eV follow a biexponential decay with time constants $\tau_2=120$ fs and $\tau_3=1.1$ ps.

In conclusion, time resolved photoemission measurements of Sr$_2$IrO$_4$ reveal the electrons dynamics upon photoexcitation above bandgap. The photoexicted holon-doublon pairs decay in to midgap states on an ultrafast timescale. Presumably, this behavior arises from emission of collective excitations and defect mediated decay. We report identical dynamics in intrinsic and doped compounds, suggesting that metallicity and screening do not influence the relaxation of the photoexcited electrons. Our time resolved photoemission data of iridates are in good agreement with optical experiments on copper oxides, providing compelling evidence of common dynamics in these intermediate Mott-Slater insulators.

We aknowledge Silke Biermann, M. Ferrero and Martin Eckstein for enlightening discussions on the physics of Mott insulators and Iridates. This work is supported by "Investissements d'Avenir" LabEx PALM (grant No. ANR-10-LABX-0039PALM), by the EU/FP7under the contract Go Fast (Grant No. 280555), and by the R\'egion Ile-deFrance through the program DIM OxyMORE.

\end{document}